\documentclass[onecollarge,natbib]{svjour2}
\bibpunct{[}{]}{;}{n}{}{,} 
\smartqed  
\usepackage{graphicx}
\usepackage{amsmath}				
\usepackage{amssymb}				
\usepackage{bm}

\def\bra#1{\mathinner{\langle{#1}|}}
\def\ket#1{\mathinner{|{#1}\rangle}}

\journalname{Few Body Systems}
\begin{document}

\title{On the microscopic structure of $\pi NN$, $\pi N\Delta$ and $\pi\Delta\Delta$
vertices}


\author{Ju-Hyun Jung \and
        Wolfgang Schweiger
}


\institute{J.-H. Jung \and W. Schweiger \at
              Institut f\"ur Physik, FB Theoretische Physik, Universit\"at Graz, Austria\\
               \email{ju.jung@uni-graz.at}\\
              \email{wolfgang.schweiger@uni-graz.at}           
}

\date{Received: date / Accepted: date}

\maketitle

\begin{abstract}
We use a hybrid constituent-quark model for the microscopic description of $\pi N N$, $\pi N \Delta$ and $\pi \Delta \Delta$ vertices. In this model quarks are confined by an instantaneous potential and are allowed to emit and absorb a pion, which is also treated as dynamical degree of freedom. The point form of relativistic quantum mechanics is employed to achieve a relativistically invariant description of this system. Starting with an $SU(6)$ spin-flavor symmetric wave function for $N_0$ and $\Delta_0$, i.e. the eigenstates of the pure confinement problem, we calculate the strength of the $\pi N_0 N_0$, $\pi N_0 \Delta_0$ and $\pi \Delta_0 \Delta_0$ couplings and the corresponding vertex form factors. Interestingly the ratios of the resulting couplings resemble strongly those needed in purely hadronic coupled-channel models, but deviate significantly from the ratios following from SU(6) spin-flavor symmetry in the non-relativistic constituent-quark model.

\end{abstract}

\section{Motivation and Formalism}
Our interest in $\pi N N$, $\pi N \Delta$ and $\pi \Delta \Delta$ couplings and vertex form factors is connected with our attempt to take pion-loop effects into account when describing the electromagnetic structure of $N$ and $\Delta$ within a constituent-quark model. As it turns out, the calculation of the loop effects boils down to a purely hadronic problem, in which the quark substructure of the $N$ and the $\Delta$ is hidden in strong and electromagnetic form factors of \lq\lq bare\rq\rq\ baryons, i.e. eigenstates of the pure confinement problem. Since $\pi N N$, $\pi N \Delta$ and $\pi \Delta \Delta$ couplings and vertex form factors are basic building blocks of nuclear physics and every hadronic model of meson-baryon dynamics, their microscopic description is also highly desirable on more fundamental grounds.

Our starting point for calculating the strong $\pi NN$, $\pi N\Delta$ and $\pi\Delta\Delta$ couplings and form factors is the mass-eigenvalue problem for three quarks that are confined by an instantaneous potential and can emit and reabsorb a pion. To describe this system in a relativistically invariant way, we make use of the point-form of relativistic quantum mechanics. Employing the Bakamjian-Thomas construction, the overall four-momentum operator $\hat{P}^{\mu}$ can be separated into a free 4-velocity operator $\hat{V}^{\mu}$ and an invariant mass operator $\hat{\mathcal{M}}$ that contains all the internal motion, i.e. $\hat{P}^{\mu}=\hat{\mathcal{M}}\,\hat{V}^{\mu}$~\cite{Biernat:2010tp}. Bakamjian-Thomas-type mass operators are most conveniently represented
by means of velocity states $\vert V;{\bf k}_{1},\mu_{1};{\bf k}_{2},\mu_{2};\dots;{\bf k}_{n},\mu_{n}\rangle$, which specify an $n$-body system by its overall velocity $V$ ($V_{\mu}V^{\mu}=1$), the CM momenta ${\bf k}_{i}$ of the individual particles and their (canonical) spin projections $\mu_{i}$~\cite{Biernat:2010tp}. Since the physical baryons of our model contain, in addition to the $3q$-component, also a $3q\pi$-component, the mass eigenvalue problem can be formulated as a 2-channel problem of the form
\begin{equation}
\left(\begin{array}{cc}
\hat{M}_{3q}^{\mathrm{conf}} & \hat{K}_{\pi}\\
\hat{K}_{\pi}^{\dag} & \hat{M}_{3q\pi}^{\mathrm{conf}}
\end{array}\right)\left(\begin{array}{c}
\ket{\psi_{3q}}\\
\ket{\psi_{3q\pi}}
\end{array}\right)=m\left(\begin{array}{c}
\ket{\psi_{3q}}\\
\ket{\psi_{3q\pi}}
\end{array}\right)\,,\label{eq:mmat}
\end{equation}
with $\ket{\psi_{3q}}$ and $\ket{\psi_{3q\pi}}$ denoting the two Fock-components of the physical baryon states $\ket{B}$. The mass operators on the diagonal contain, in addition to the relativistic particle energies, an instantaneous confinement potential between the quarks. The vertex operator $\hat{K}_{\pi}^{(\dag)}$ connects the two channels and describes the absorption (emission) of the $\pi$ by one of the quarks. Its velocity-state representation can be directly connected to a corresponding field-theoretical interaction Lagrangean~\cite{Biernat:2010tp}. We use a pseudovector interaction Lagrangean for the $\pi qq$-coupling
\vspace{-0.1cm}
\begin{equation}
\mathcal{L}_{\pi qq}(x)=-\frac{f_{\pi qq}}{m_{\pi}}\left(\bar{\psi}_{q}(x)\gamma_{\mu}\gamma_{5}\vec{\tau}\psi_{q}(x)\right)
\cdot\partial^{\mu}\vec{\phi}_{\pi}(x)\label{eq:piqqvertex}\, .
\end{equation}

\vspace{-0.1cm}
\noindent After elimination of the $3q\pi$-channel the mass-eigenvalue equation takes on the form
\begin{equation}
\bigl[\hat{M}_{3q}^{\mathrm{conf}}+\underbrace{\hat{K}_{\pi}(m-\hat{M}_{3q\pi}^{\mathrm{conf}})^{-1}
\hat{K}_{\pi}^{\dag}}_{\hat{V}_{\pi}^{\mathrm{opt}}(m)}\bigr]\ket{\psi_{3q}}=m\,\ket{\psi_{3q}}\,,
\label{eq:Mphys}
\end{equation}

\vspace{-0.2cm}
\noindent
where $\hat{V}_{\pi}^{\mathrm{opt}}(m)$ is an optical potential that describes the emission and reabsorption of the pion by the quarks. One can now solve Eq.~(\ref{eq:Mphys}) by expanding the ($3q$-components of the) eigenstates in terms of eigenstates of the pure confinement problem, i.e. $\ket{\psi_{3q}}=\sum_{B_{0}}\alpha_{B_{0}}\,\ket{B_{0}}$, and determining the open coefficients $\alpha_{B_{0}}$. Since the particles that propagate within the pion loop are also bare baryons (rather than quarks), the problem of solving the mass eigenvalue equation~(\ref{eq:Mphys}) reduces then to a pure hadronic problem, in which the dressing and mixing of bare baryons by means of pion loops produces finally the physical baryons. The quark substructure determines just the coupling strengths at the pion-baryon vertices and leads to vertex form factors. To set up the mass-eigenvalue equation on the hadronic level one needs matrix elements $\bra{B_{0}^{\prime}}\hat{V}_{\pi}^{\mathrm{opt}}(m)\ket{B_{0}}$ of the optical potential between bare baryon states. The general structure of these matrix elements is ($B_{0}$ and $B_{0}^{\prime}$ are at rest)
\begin{equation}
\bra{B_{0}^{\prime}}\hat{V}_{\pi}^{\mathrm{opt}}(m)\ket{B_{0}}\propto\sum_{B_{0}^{\prime\prime}}
\int\frac{d^{3}k_{\pi}^{\prime\prime}}{\sqrt{m_{\pi}^{2}+{\vec{k}_{\pi}^{\prime\prime\,^{2}}}} \sqrt{m_{B_0^{\prime\prime}}^{2}+{\vec{k}_{\pi}^{\prime\prime\,^{2}}}}}\,
J_{\pi B_{0}^{\prime} B_{0}^{\prime\prime}}^{5} (\vec{k}_{\pi}^{\prime\prime})\,\frac{1}{m-m_{B_{0}^{\prime\prime}\pi}}\,J_{\pi B_{0}^{\prime\prime}B_{0}}^{5}(\vec{k}_{\pi}^{\prime\prime})\ ,\label{eq:vopt}
\end{equation}
where $m_{B_{0}^{\prime\prime}\pi}$ is the invariant mass of the $B_{0}^{\prime\prime}\pi$ system in the intermediate state and spin- as well as isospin dependencies have been suppressed.

For the cases we are interested in, i.e. the $N$ and the $\Delta$, the currents occurring in Eq.~(\ref{eq:vopt}) can be cast into the form\footnote{This form exhibits the correct chiral properties and avoids problems with superfluous spin degrees of freedom when treating spin-3/2 fields covariantly by means of Rarita-Schwinger spinors~\cite{Pascalutsa:2005nd}.}:
\vspace{-0.2cm}
\begin{eqnarray}
J_{\pi N_{0}N_{0}}^{5}(\vec{k}_{\pi}) & = & i\,\frac{f_{\pi N_{0}N_{0}}}{m_{\pi}}F_{\pi N_{0}N_{0}}(\vec{k}_{\pi}^{2})\,\bar{u}(-\vec{k}_{\pi})\gamma_{\mu}\gamma_{5}u(\vec{0})\,k_{\pi}^{\mu}\,,\nonumber \\
J_{\pi\Delta_{0}\Delta_{0}}^{5}(\vec{k}_{\pi}) & = & \frac{f_{\pi\Delta_{0}\Delta_{0}}}{m_{\pi}m_{\Delta_{0}}}F_{\pi\Delta_{0}\Delta_{0}}(\vec{k}_{\pi}^{2})\,\epsilon^{\mu\nu\rho\sigma}\,\bar{u}_{\mu}(-\vec{k}_{\pi})\,u_{\nu}(\vec{0})\,k_{\Delta_{0},\rho}\,k_{\pi,\sigma}\,,\nonumber \\
J_{\pi N_{0}\Delta_{0}}^{5}(\vec{k}_{\pi}) & = & -i\frac{f_{\pi N_{0}\Delta_{0}}}{m_{\pi}m_{\Delta_{0}}}\,F_{\pi N_{0}\Delta_{0}}(\vec{k}_{\pi}^{2})\,\epsilon^{\mu\nu\rho\sigma}\,\bar{u}(-\vec{k}_{\pi})\gamma_{\sigma}\gamma_{5}u_{\nu}(\vec{0})\,k_{\Delta_{0},\mu}\,k_{\pi,\rho}\,,\nonumber \\
J_{\pi\Delta_{0}N_{0}}^{5}(\vec{k}_{\pi}) & = & i\frac{f_{\pi N_{0}\Delta_{0}}}{m_{\pi}m_{\Delta_{0}}}\,F_{\pi\Delta_{0}N_{0}}(\vec{k}_{\pi}^{2})\,\epsilon^{\mu\nu\rho\sigma}\,\bar{u}_{\nu}(-\vec{k}_{\pi})\gamma_{5}\gamma_{\sigma}u(\vec{0})\,k_{\Delta_{0},\mu}\,k_{\pi,\rho}\,, \label{eq:currents}
\end{eqnarray}

\vspace{-0.1cm}
\noindent where $u(.)$ is the Dirac spinor of the nucleon and $u_{\mu}(.)$ the Rarita-Schwinger spinor of the $\Delta$. 
Here we have again suppressed the isospin dependence and also omitted the spin labels.
From Eqs.~(\ref{eq:vopt}) and (\ref{eq:currents}) one can then infer the analytical expression
for the combination $f_{\pi B_{0}^{\prime}B_{0}}\,F_{\pi B_{0}^{\prime}B_{0}}(\vec{k}_{\pi}^{2})$ in terms of quark degrees of freedom~\cite{Kupelwieser:2016}.

Assuming a scalar isoscalar confinement potential, the masses of the bare nucleon and the bare Delta are degenerate, the momentum part of the wave function will be the same and the spin-flavor part of the wave function is $SU(6)$ symmetric. Rather than solving the confinement problem for a particular potential, we parameterize the momentum part of the $3q$ wave function of $N_{0}$ and $\Delta_{0}$ by means of a Gaussian
\begin{equation}
\psi_{3q}^{N_{0},\Delta_{0}}(\vec{k}_{q_{1}},\vec{k}_{q_{2}},\vec{k}_{q_{3}})\propto
\exp\left(-\alpha^{2}(\vec{k}_{q_{1}}^{2}+\vec{k}_{q_{2}}^{2}+\vec{k}_{q_{3}}^{2})\right)\,,
\quad\vec{k}_{q_{1}}+\vec{k}_{q_{2}}+\vec{k}_{q_{3}}=\vec{0}\,, \label{eq:3qwf}
\end{equation}
and consider the mass of $N_{0}$ and $\Delta_{0}$, i.e. $M_{N_{0}}=M_{\Delta_{0}}=:M_{0}$ as free parameter. The parameters of our model
are therefore the oscillator parameter $\alpha$, the mass $M_{0}$, the constituent-quark mass $m_{q}:=m_{u}=m_{d}$
and $f_{\pi qq}$, the $\pi qq$ coupling strength. For fixed $m_{q}=263$~MeV (taken from Ref.~\cite{Pasquini:2007iz}) we have adapted the remaining parameters such that the physical $N$ and $\Delta$ masses, resulting from the mass renormalization due to pion loops (with $N_{0}$ and $\Delta_{0}$ intermediate states), agree with their experimental values. In order to tune these parameters we started with a fixed $M_0$ and $\alpha$ and calculated the masses of the physical nucleon and Delta as function of $f_{\pi q q}$ by solving the mass eigenvalue equation~(\ref{eq:Mphys}). Then we have varied $M_0$ and $\alpha$ such that the physical nucleon and Delta masses (i.e. $m_N=0.9385$~GeV and $m_\Delta=1.233$~GeV) are obtained for a reasonable value of $f_{\pi q q}$ (which also leads to a reasonable value for $f_{\pi N_0 N_0}$). This gives us $M_{0}=1.350$~GeV, $\alpha=2.915$~GeV$^{-1}$ and $f_{\pi qq}=0.6602$. Note that in our simple model the physical $\Delta$ is still a stable particle, since the threshold of the only possible decay channel $\pi N_0$ is larger than the mass of the physical $\Delta$. In order to get a $\Delta$ with a finite decay width within our model one would need in Eq.~(\ref{eq:mmat}) an additional $3q\pi\pi$ channel. For a purely hadronic model which provides a finite decay width for the $\Delta$ and makes use of the same relativistic coupled-channel framework as employed here, see Ref.~\cite{Plessas2016}.
\vspace{-0.3cm}

\section{Results and Discussion}
\vspace{-0.2cm}
\begin{figure}[t!]
\includegraphics[width=0.48\textwidth]{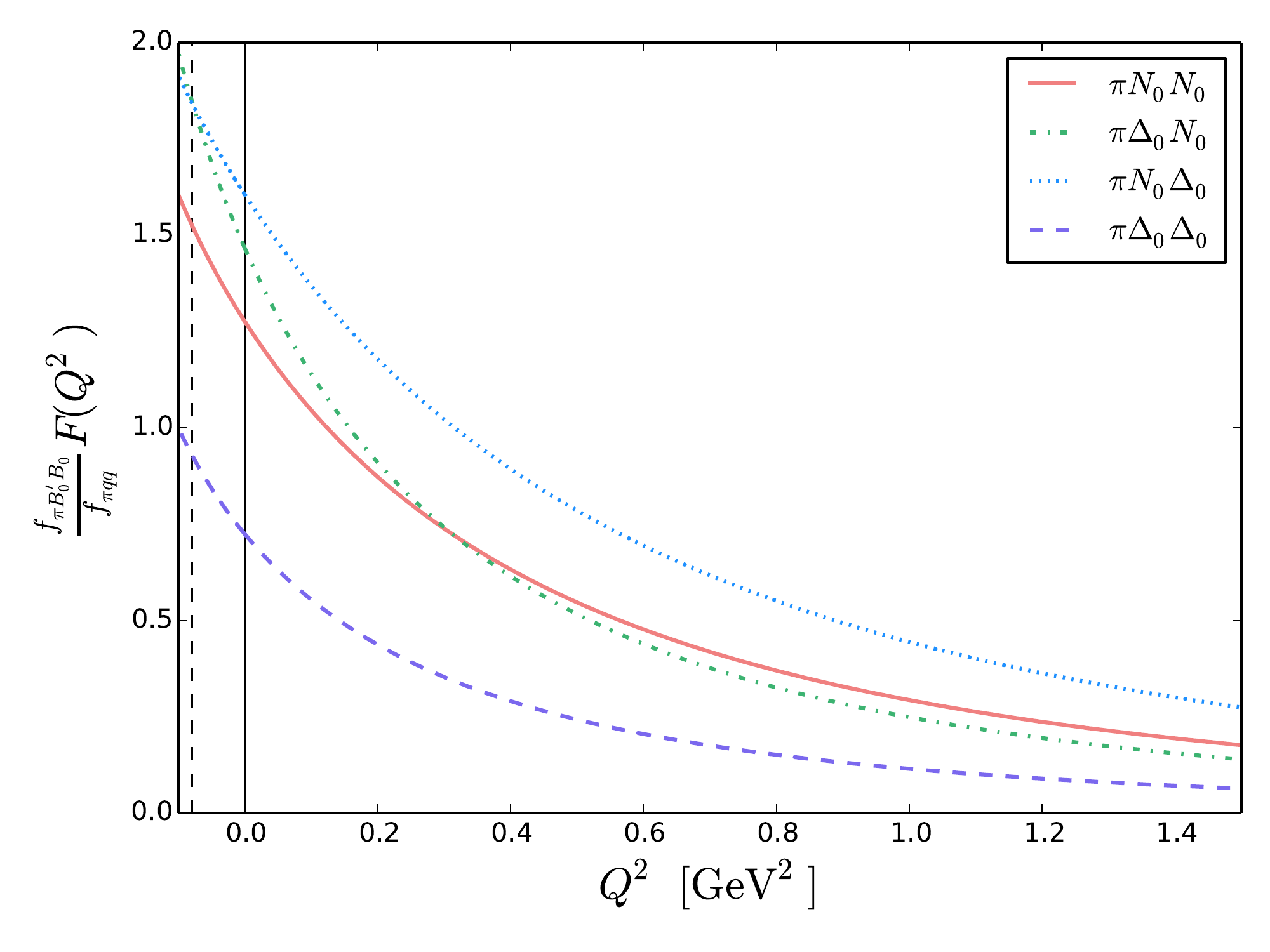}\hfill\includegraphics[width=0.48\textwidth]{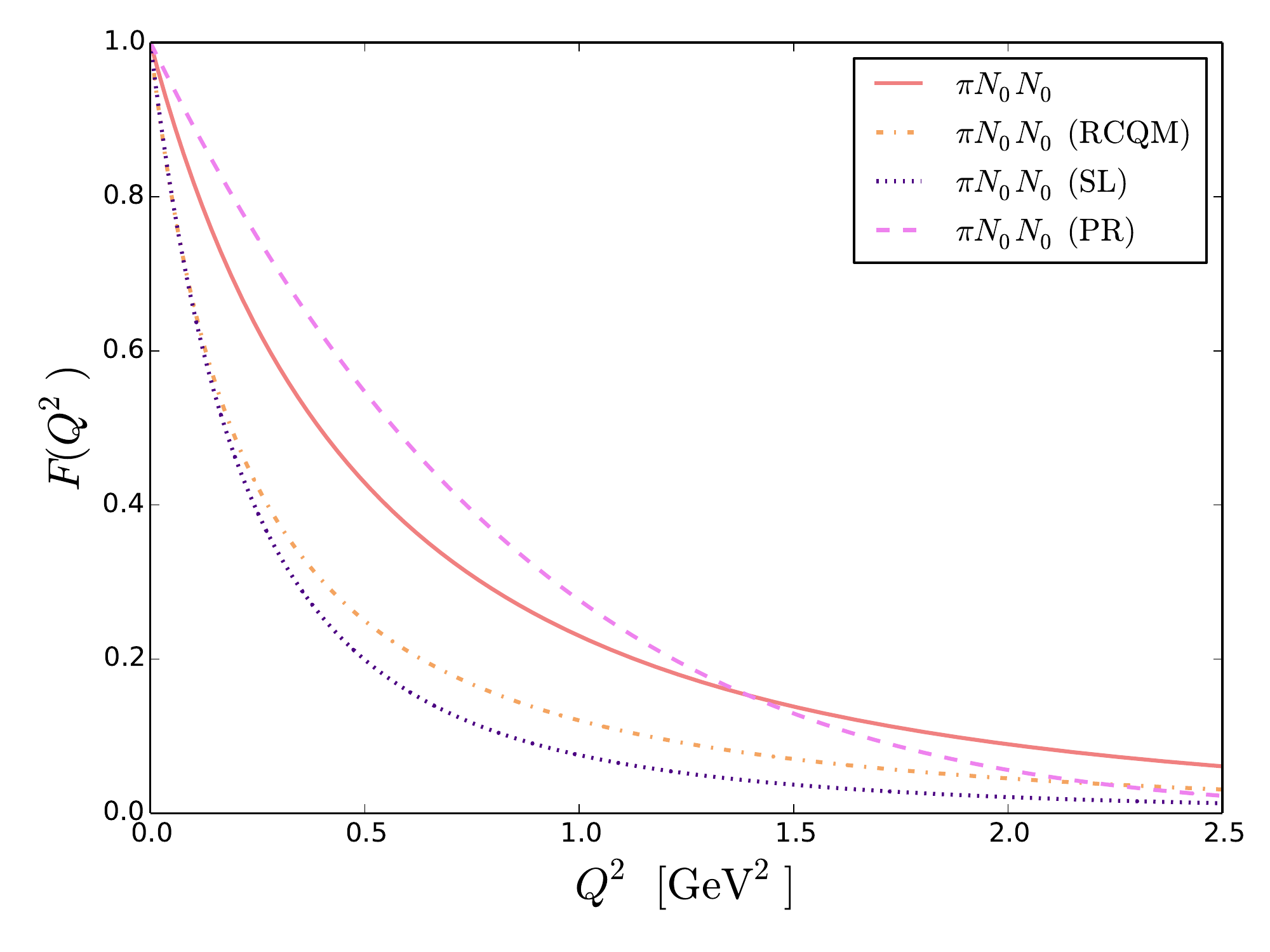}
\vspace{-0.3cm}
\caption{The left plot shows the (unnormalized) $\pi N_{0}N_{0}$, $\pi\Delta_{0}\Delta_{0}$,
$\pi N_{0}\Delta_{0}$, and $\pi\Delta_{0}N_{0}$ form factors as functions of $Q^{2}=-2M_{0}(M_{0}-(M_{0}^{2}+\vec{k}_{\pi}^{2})^{1/2})$.
In the right plot the $Q^{2}$ behavior of $F_{\pi N_{0}N_{0}}$ (normalized to $1$ at $Q^{2}=0$) is compared to the outcome of another relativistic constituent-quark model~\cite{Melde:2008dg} and to phenomenological fits obtained within two purely hadronic dynamical coupled-channel models~\cite{Kamano:2013iva,Polinder:2005sn} (SL
and PR).}\label{fig:ffs}\vspace{-0.5cm}
\end{figure}
The left plot of Fig.~\ref{fig:ffs} shows (unnormalized)
$\pi N_{0}N_{0}$, $\pi\Delta_{0}\Delta_{0}$,
$\pi N_{0}\Delta_{0}$, and $\pi\Delta_{0}N_{0}$
form factors as function of the (negative) four-momentum transfer
squared (analytically continued to small time-like momentum transfers).
It is worth noting that $F_{\pi\Delta_{0}N_{0}}$ and $F_{\pi N_{0}\Delta_{0}}$
do not coincide. This is, of course, no surprise, since in the first
case the $N_{0}$ is real and the $\Delta_{0}$ virtual, whereas it
is just the other way round in the second case. The form factors describe
thus completely different kinematical situations, but they coincide
at a particular negative (i.e. unphysical) value of $Q^{2}$. Since
there is only one coupling strength at the $\pi N_{0}\Delta_{0}$-vertex
(i.e. $f_{\pi\Delta_{0}N_{0}}=f_{\pi N_{0}\Delta_{0}}$, see Eq.~(\ref{eq:currents})),
this is the natural point to normalize the form factors and extract
the coupling constants. Its value $Q_{0}^{2}=-0.079$~GeV$^{2}$ is close
to the standard normalization point, namely the pion pole $Q_{0}^{2}=-m_{\pi}^{2}$.
The resulting coupling strengths are given in Tab.~\ref{tab:couplings} and compared with those from other models.

The couplings quoted for the hadronic model~\cite{Polinder:2005sn} are the physical (\lq\lq dressed\rq\rq) couplings. The couplings given in Ref.~\cite{Melde:2008dg} may be interpreted as \lq\lq bare\rq\rq\ couplings, since meson-cloud effects are not included explicitly, but physical $N$ and $\Delta$ masses are used for their extraction. The values for $f_{\pi N \Delta}$ and $f_{\pi \Delta\Delta}$ taken from Ref.~\cite{Kamano:2013iva} are bare couplings. Dressing of the nucleon, however, is not considered by the authors. Our results are also bare couplings, since they have been extracted from vertex matrix elements $\langle B_0^\prime \pi |K_\pi^\dag |B_0\rangle$ involving only bare baryons.
For the ratio of the coupling strengths we get $f_{\pi N_{0}\Delta_{0}}:f_{\pi N_{0}N_{0}}:f_{\pi\Delta_{0}\Delta_{0}}=1.209:1:0.608$.
This may be compared with the prediction from the non-relativistic constituent-quark model
assuming $SU(6)$ spin-flavor symmetry, i.e. $f_{\pi N\Delta}:f_{\pi NN}:f_{\pi\Delta\Delta}=4\sqrt{2}/5:1:9/5=1.13:1:1.8$~\cite{Brown:1975di}.
The differences can solely be ascribed to relativistic effects and
are obviously significant, in particular for the $\pi\Delta_{0}\Delta_{0}$-vertex.
Remarkably, our fractions resemble very much those needed in the dynamical
coupled-channel model of Ref.~\cite{Kamano:2013iva}, i.e. $f_{\pi N\Delta}:f_{\pi NN}:f_{\pi\Delta\Delta}=1.26:1:0.42$.

In the right plot of Fig.~\ref{fig:ffs} our result for $F_{\pi N_{0}N_{0}}$
is compared with the outcome of another relativistic constituent-quark
model~\cite{Melde:2008dg} and with two parameterizations of this
form factor that have been used in the hadronic models~\cite{Kamano:2013iva,Polinder:2005sn}.
Up to $Q^{2}\approx1$~GeV$^{2}$ our prediction is comparable with
the form factor parametrization of Ref.~\cite{Polinder:2005sn},
but for higher $Q^{2}$ it falls off slower\footnote{It is, of course, an extreme point of view to attribute the $N$-$\Delta$ mass difference solely to the pions. As a consequence, the $3q$ wave function in Eq.~(\ref{eq:3qwf}) and thus also strong and electromagnetic form factors may fall off too slowly. If this will turn out in our calculations of the electromagnetic form factors one has to think of including other mechanisms (like one-gluon exchange) that can also contribute to the $N$-$\Delta$ mass difference.}. The form factors of Refs.~\cite{Melde:2008dg,Kamano:2013iva}
fall off much faster already at small $Q^{2}$. Deviations of our
result from the one of Ref.~\cite{Melde:2008dg} have their origin
in different $3q$ wave functions of the nucleon, but also in different
kinematical and spin-rotation factors entering the microscopic expression
for the pseudovector current of the nucleon.

Having determined the $\pi N_{0}N_{0}$, $\pi\Delta_{0}\Delta_{0}$
and $\pi N_{0}\Delta_{0}$ vertices from a microscopic model, we are
now in the position to calculate the electromagnetic form factors
of physical nucleons and Deltas and determine the effect of pions
on their electromagnetic structure. Such calculations exist in the literature (see, e.g., Ref.~\cite{Pasquini:2007iz}), but in all the work known to us the strong pion-baryon vertices were parameterized. We rather try to treat strong and electromagnetic vertices on the same footing, i.e. start with the same three-quark wave function for the (bare) baryons, calculate strong and electromagnetic form factors of bare baryons and with these form factors as input the effect of pion loops (for physical baryons). First exploratory calculations for the nucleon (with another three-quark wave function and without Deltas) gave reasonable results and show that significant pion-loop effects can be observed for $Q^{2}\lesssim0.5$~GeV$^{2}$~\cite{Kupelwieser:2016}. It will, of course, be more interesting to investigate electromagnetic $\Delta$ and $N\rightarrow\Delta$ transition form factors, where
pionic effect are expected to play a more significant role.
\begin{table}[t!]
\caption{Our prediction for $\pi N_0 N_0$, $\pi N_0 \Delta_0$ and $\pi \Delta_0 \Delta_0$ coupling constants in comparison with results from another relativistic constituent-quark model~\cite{Melde:2008dg}. Shown are also values for these coupling constants used in two popular hadronic coupled-channel models~\cite{Kamano:2013iva,Polinder:2005sn}. The $\pi q q$ couplings employed in the quark models are given in the second column.
}\label{tab:couplings}\vspace{-0.3cm}
\begin{tabular}{|c|c|c|c|c|}
\multicolumn{1}{c}{} &  \multicolumn{1}{c}{} & \multicolumn{1}{c}{} & \multicolumn{1}{c}{}\tabularnewline
\hline
\hline
 & $f_{\pi qq}$  &  $f_{\pi {N}_{0} N_{0}}$ & $f_{\pi N_{0} \Delta_{0}}$  & $f_{\pi \Delta_{0} \Delta_{0}}$\tabularnewline
 &  &  &  & \tabularnewline
\hline
Our model & 0.6602 & 1.0027 &  1.2123 & 0.6097\tabularnewline
\hline
Ref.~\cite{Melde:2008dg} & 0.5889 &  0.9318 & 1.537 & \tabularnewline
\hline
Ref.~\cite{Kamano:2013iva} &  & 1.0027 & 1.256 & 0.415\tabularnewline
\hline
Ref.~\cite{Polinder:2005sn} &  & 0.9708 & 2.451 & \tabularnewline
\hline
\hline
\multicolumn{1}{c}{} &  \multicolumn{1}{c}{} & \multicolumn{1}{c}{} & \multicolumn{1}{c}{}\tabularnewline
\end{tabular}\vspace{-0.8cm}
\end{table}

\vspace{-0.3cm}

\begin{acknowledgements}
J.-H. Jung acknowledges the support of the Fonds zur F\"orderung der wissenschaftlichen Forschung in \"Osterreich (Grant No. FWF DK W1203-N16). \end{acknowledgements}

\vspace{-0.8cm}

\end{document}